# A HIGH RESOLUTION CAVITY BPM FOR THE CLIC TEST FACILITY*

N. Chritin, H. Schmickler, L. Soby, CERN, Geneva, Switzerland
A. Lunin, N. Solyak, M. Wendt#, V. Yakovlev, Fermilab, Batavia, IL 60510, U.S.A.

*Abstract*

In frame of the development of a high resolution BPM system for the CLIC Main Linac we present the design of a cavity BPM prototype. It consists of a waveguide loaded dipole mode resonator and a monopole mode reference cavity, both operating at 15 GHz, to be compatible with the bunch frequencies at the CLIC Test Facility. Requirements, design concept, numerical analysis, and practical considerations are discussed.

## INTRODUCTION

The next high energy physics (HEP) lepton collider, operating at the energy frontier (CM > 500 GeV) needs to accelerate the particles along a straight line, to avoid too high losses caused by synchrotron radiation. For both concepts, the superconducting RF technology based International Linear Collider (ILC) [1], as well as for the two-beam acceleration Compact LInear Collider (CLIC) [2], generation and preservation of beams with ultra-low transverse emittance is mandatory to achieve the luminosity goals. A “golden” trajectory path has to be established, steering the beam with sub-micrometer precision through the magnetic centres of the quadrupoles, avoiding a beam blow-up due to non-linear fields.

Table 1: CLIC / CTF Main Linac BPM specifications

| | **CLIC** | **CTF** CALIFES |
|---|---|---|
| Nominal bunch charge [nC] | 0.6 | 0.6 |
| Bunch length (RMS) [μm] | 44 | 225 |
| Batch length [nsec] | 156 | 1-150 |
| Bunch spacing [nsec] | 0.5 | 0.6667 |
| Beam pipe radius [mm] | 4 | 4 |
| BPM time resolution [nsec] | <50 | <50 |
| BPM spatial resolution [nm] | <50 | <50 |
| BPM stability [nm] | <100 | <100 |
| BPM accuracy [μm] | <5 | <5 |
| BPM dynamic range [μm] | ±100 | ±100 |
| BPM resonator frequency [GHz] | 14 | 14.98962 |

To discover and minimize unwanted dispersion effects along the Main Linac beam-line, an energy chirp modulation within the bunch train is proposed for the CLIC beam commissioning. Therefore the beam position monitors (BPM) have to have both, high spatial and high time resolution. We propose a low-Q waveguide loaded $TM_{110}$ dipole mode cavity as BPM, which is complemented by a $TM_{010}$ monopole mode resonator of same resonant frequency for reference signal purposes.

Table 1 gives the basic requirements for the BPM, however, as the bunch frequencies at the CLIC Test Facility (CTF) are different from the proposed CLIC linear collider, the nominal resonant frequencies ($f_{110}$ for the dipole mode cavity, $f_{010}$ for the reference cavity) are not the same for the CTF prototype BPM, and the CLIC Main Linac BPMs:

- CLIC: 14 GHz
- CTF: 15 GHz

Choosing a rather high operating frequency $n\ f_{\mathrm{bunch}}$ has several advantages, e.g. most higher-order modes (HOM) are damped by the beam pipe cut-off frequency, and higher shunt impedances can be achieved (better sensitivity, higher resolution potential). However, as dipole mode and reference cavity operate at the same frequency, we have to ensure that they do not couple by evanescence fields leaking into the beam pipe.

With a time resolution of <50 nsec we will be able to acquire three beam position samples within the 156 nsec long bunch train (batch). Because of dynamic range limitations in the read-out system, the high 50 nm spatial resolution can only be accomplished within a small range of ±100 μm beam displacement from the BPM center, however a moderate resolution (some μm) will be achievable over the full aperture (±4 mm).

Each of the two ~20 km long CLIC Main Linacs will be equipped with ~2000 BPMs, therefore a simple, cost-effective design with minimum maintenance and calibration requirements is important. The longitudinal expansion of the two BPM resonators should not exceed 95 mm, including flanges. A simple analog down-converter has to be placed in close proximity to the BPM detector; no external reference or timing signals are required.

## CAVITY BPM PRINCIPLE

A cylindrical “pillbox” having conductive (metal) walls of dimensions radius $R$ and length $l$ resonates at eigenfrequencies:

$$f_{\mathrm{mnp}} = \frac{1}{2\pi\sqrt{\mu_0 \varepsilon_0}} \sqrt{\left(\frac{j_{mn}}{R}\right)^2 + \left(\frac{p\pi}{l}\right)^2} \qquad (1)$$

This resonator can be utilized as passive, beam driven cavity BPM by assembling it into the vacuum beam pipe. A subset of these eigenmodes is excited by the bunched beam, for use as BPM the lowest transverse-magnetic dipole mode $TM_{110}$ is of interest ($n\, f_{\text{bunch}} = f_{110}$). Its

$$E_z = C\, J_1 \frac{j_{11} r}{R} \cos\phi\, e^{i\omega} \qquad (2)$$

field component couples to the beam, with almost linear dependence to the beam displacement $r$, and beam intensity (hidden in the constant $C$).

*Work supported by the Fermi National Accelerator laboratory, operated by Fermi Research Alliance LLC, under contract No. DE-AC02-07CH11359 with the US Department of Energy.
#manfred@fnal.gov

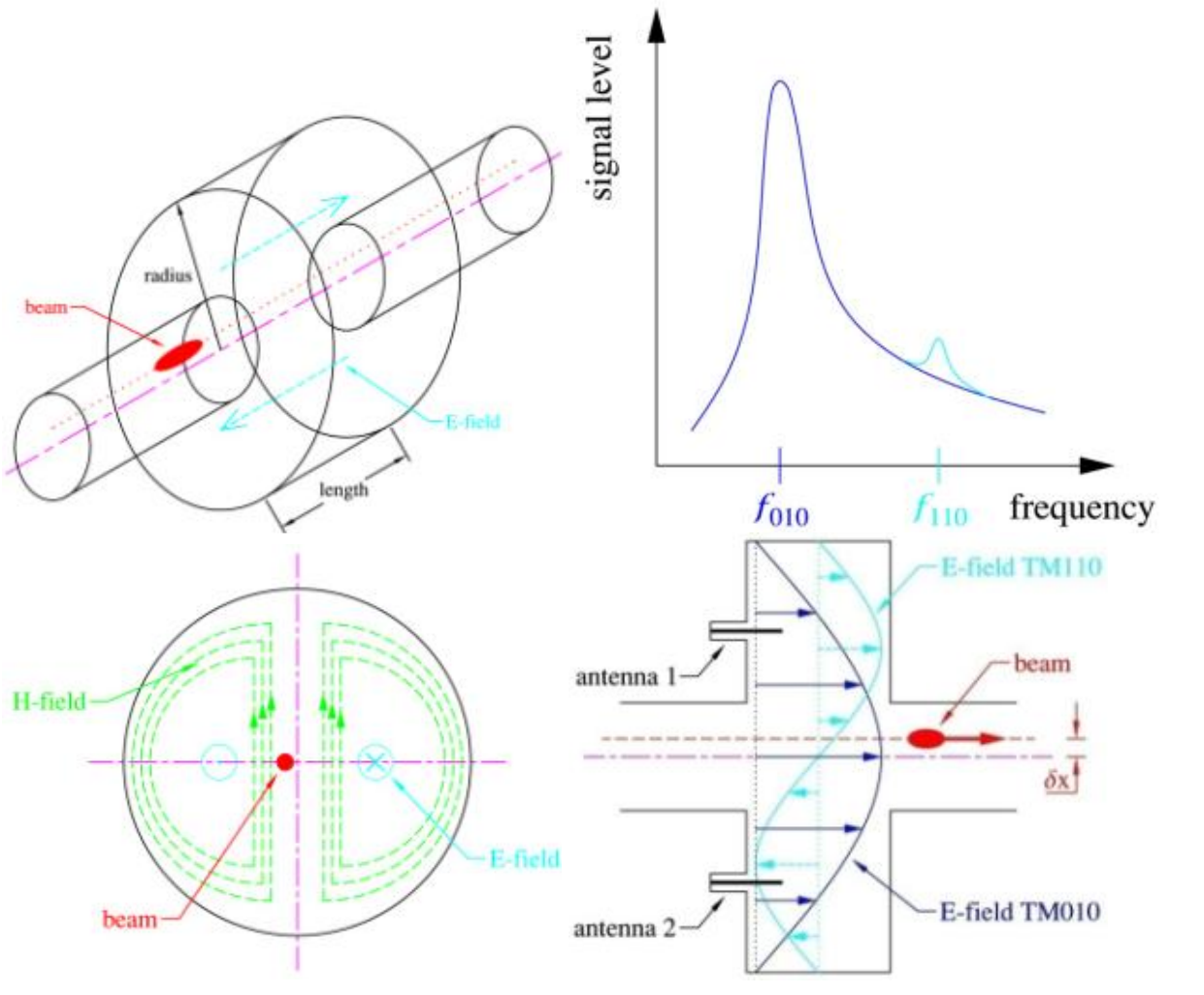

Figure 1: Cavity BPM principle.

As Fig. 1 illustrates, the $TM_{110}$ dipole mode vanishes as the beam moves exactly through the center of the cavity pickup ($r$ = 0). A set of capacitive coupling pin-antennas or other coupling schemas can be used to sense the dipole mode displacement signal of frequency $f_{110}$. Compared to broadband BPM pickups, e.g. button- or stripline style, the strength of the beam excited eigenmode is due to the cavity shape. For a simple cylindrical cavity BPM the corresponding $R/Q$ (shunt impedance / quality factor) typically has much higher values when compared to the transfer impedance of a broadband BPM, one achieves a higher beam displacement sensitivity and therefore a greater potential to operate as "high resolution" BPM.

## "CM-FREE", LOW-Q CLIC BPM

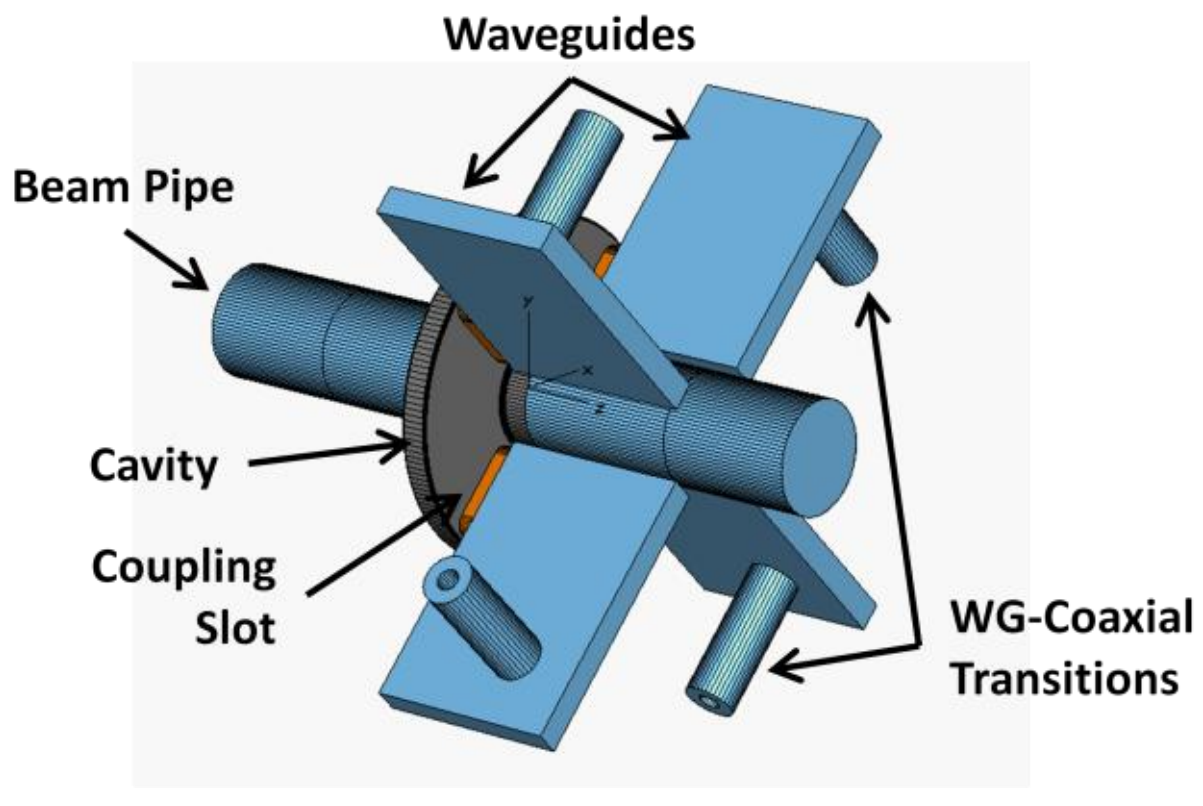

Figure 2: CLIC cavity BPM.

Figure 2 shows the proposed CLIC cavity BPM. A set of four, symmetrically arranged waveguides are slot-coupled to the dipole mode resonator. The $TE_{10}$ cut-off frequency of the rectangular waveguide is choosen between the first monopole and dipole mode of the cavity,

$$f_{TM010(Cav)} < f_{\text{TE10(WG)}} = \frac{1}{2a\sqrt{\varepsilon\mu}} < f_{TM110(Cav)} \qquad (3)$$

and acts like a high-pass filter to suppress the strong monopole mode signal. The four narrow slots between cavity and waveguides sets the $TM_{110}$ polarization axes in the resonator to the horizontal, respectively vertical plane, and provide magnetic coupling of the beam displacement signal to the waveguides. A feedthrough pin antenna is used for the waveguide-to-coaxial transition of 50 Ω characteristic impedance. Figure 3 illustrates an *Ansoft* HFSS simulation, showing the broadband behaviour of the WG-coaxial transition, BW~1 GHz for -30 dB return loss. We plan to read-out all four coaxial ports, using a hybrid combiner may further improve the suppression of higher-order modes (phase filtering). We also can utilize the 2$^{\text{nd}}$ ports for calibration and redundancy purposes.

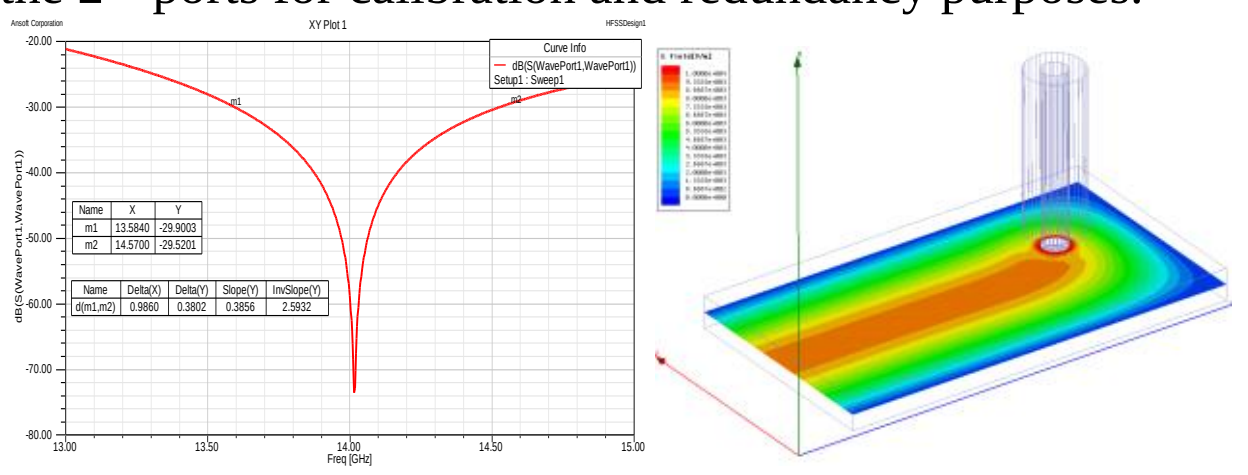

Figure 3: Matching of the WG-to-coaxial transition.

When excited with a beam bunch, the resonant BPM cavity behaves like a damped oscillator, with an impulse response voltage signal:

$$\text{v}(t) \propto \text{v}_0\, e^{-t/\tau} \cos(\omega_0 t) \qquad (4)$$

where $\tau = 2Q_l/\omega_0$ is the resonator time constant, and $Q_l$ is the loaded quality factor. The required BPM time resolution (<50 nsec) and dynamic range (±100μm) limits the maximum loaded quality factor of the cavity. The amplitude of the $TM_{110}$ dipole mode should decay $10^3$ times within 50 nsec in order to achieve the required resolution. Thus, the maximum loaded Q-factor is given by:

$$Q_{\max} = \frac{t_{\max}\,\omega_0}{2\ln(1000)} \approx 320 \qquad (5)$$

Keeping the cavity dimensions reasonable and applying the waveguide load impedance is not sufficient to achieve this rather low Q-value, thus the resonator has to be manufactured out of a more lossy material, i.e. stainless steel.

## BEAM SPECTRUM ESTIMATION

The BPM cavity, with its four slot-coupled waveguides, terminated by WG-coaxial transition ports, and beam pipe ports is a complex resonant system. A single beam bunch will excite most eigenmodes, from which only the $TM_{110}$ cavity dipole mode is of interest for our beam position measurement, Figure 4 illustrates the coupling mechanism.

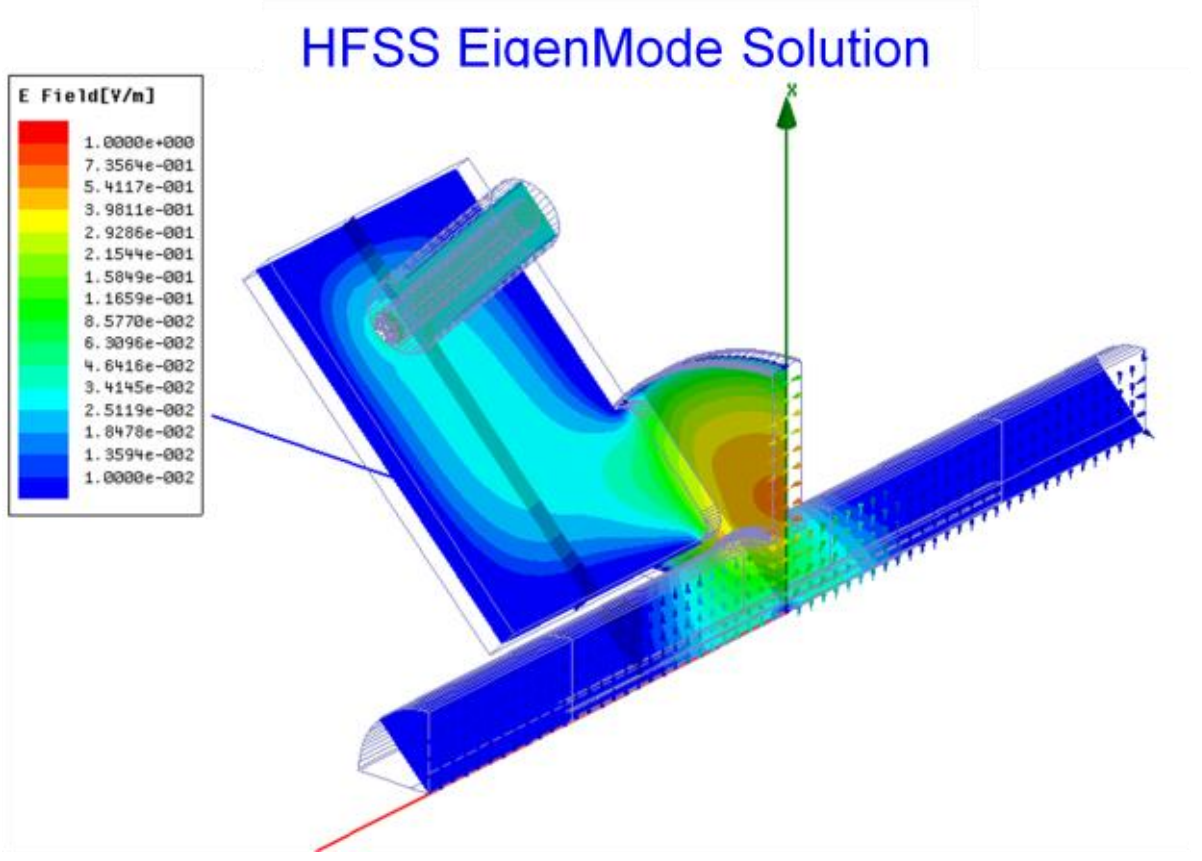

Figure 4: E-field of the $TM_{110}$ eigenmode.

However, unwanted modes, even though they have different resonance frequencies, will perturb the $TM_{110}$ dipole mode because their energy is spread over a wide range of frequencies, due to their limited Q-value (mode leakage). For each mode below the beam pipe cut-off frequency (22 GHz) we estimated the signal voltage (at t=0) at the coaxial output port:

$$\mathrm{v} = q\,\omega\sqrt{\frac{Z_0\,(R/Q)}{Q_{ext}}} \qquad (6)$$

where $q$ is the bunch charge, $Z_0$=50 Ω is the characteristic impedance of the coaxial port, and $R/Q$ is the characteristic impedance of the cavity for a specific mode. In (6) the signals of all coaxial ports are summed, to report monopole mode signals, $V_{0mp}$ has to be divided by 2, for the dipole modes divide $V_{1mp}$ by √2. Table 2 lists the results for the relevant eigenmodes.

The computed output voltage for each mode (CLIC BPM frequencies) in Table 2 has to be multiplied by the expected rejection coefficients, i.e. frequency selectivity, phase filtering and multi-bunch rejection. The frequency filter rejection $KF_{nmp}$ is simply the normalized signal intensity of an unwanted mode $TM_{nmp}$ leaking into the $TM_{110}$ dipole mode (at $f_{110}$):

$$\mathrm{KF}_{nmp} = \frac{\mathrm{V}_{nmp}(\omega_{110})}{\mathrm{V}_{nmp}(\omega_{nmp})} \qquad (7)$$

By summing the two opposite port signals we can benefit in additional ~20 dB (0.1) phase filter rejection for the $TM_{210}$ mode.

In the practical application, the cavity BPM will not be excited by a single bunch, but by a train of typically $k$=312 bunches, spaced by 0.5 nsec. Therefore the output signal is a superposition of the single bunch responses of all modes, assuming the same intensity of all bunches:

$$\mathrm{V}_{sum}(t,k) = \sum_{1}^{k} \mathrm{V}_{010}\, e^{-\beta_{010}(t-t_b k)} \cos[\omega_{010}(t-t_b k)] \qquad (8)$$

$$+\sum_{1}^{k} \mathrm{V}_{110}\, e^{-\beta_{110}(t-t_b k)} \cos[\omega_{110}(t-t_b k)] + \ldots$$

While $f_{110}$ is in phase with the bunch frequency$1/t_b$, resulting in a positive signal pile-up, the unwanted modes receive a “random” multi-bunch excitation, which leads to a rejection with respect to the beam position signal:

$$\mathrm{KM}_{010} = \frac{\int_0^{t_{max}} \sum_{1}^{k} \mathrm{V}_{110}\, e^{-\beta_{110}(t-t_b k)} \cos[\omega_{110}(t-t_b k)]}{\int_0^{t_{max}} \sum_{1}^{k} \mathrm{V}_{010}\, e^{-\beta_{010}(t-t_b k)} \cos[\omega_{010}(t-t_b k)]} \qquad (9)$$

Table 2: Results of the modal analysis of the CLIC cavity BPM

| **Mode Type** | | **Freq.** [GHz] | $Q_{tot}$[1], ($Q_l$) | ***R/Q*** [Ω], [Ω/mm$^2$], [Ω/mm$^4$] | **Output Voltage**[2,3] [V], [V/mm], [V/mm$^2$] | **Frequency Filter Rejection** | **Phase Filter Rejection**[4] | **Multi-bunch Regime Rejection** |
|---|---|---|---|---|---|---|---|---|
| $TM_{010}$ | | 10.385 | 380, (>$10^9$) | 45 | <0.001 | 0.005 | - | 0.1 |
| $TM_{110}$ | | 13.999 | 250, (540) | 3 | 17 | - | - | - |
| $TM_{210}$ | | 18.465 | 80, (100) | 0.05 | 5 | 0.025 | 0.1 | 0.1 |
| $TM_{020}$ | | 24.300 | 680, (>$10^9$) | 12 | <0.001 | 0.001 | - | 0.05 |
| WG1 | $TM_{11}$ | 12.285 | 6 | - | 3 | - | - | - |
| | $TM_{21}$ | 12.285 | 6 | - | 0.3 | - | - | - |
| WG2 | $TM_{11}$ | 15.878 | 4 | - | 5 | - | - | - |
| | $TM_{21}$ | 15.880 | 4 | - | 1.2 | - | - | - |
| WG3 | $TM_{21}$ | 21.610 | 7 | - | - | - | - | - |

[1] - Stainless steel resonator material
[2] – RMS value; normalized to 1 nC charge
[3] - Signals are from a single coaxial output at the eigenmode frequency. Multipole modes are normalized to 1 mm off-axis shift
[4] – For $TM_{210}$ only

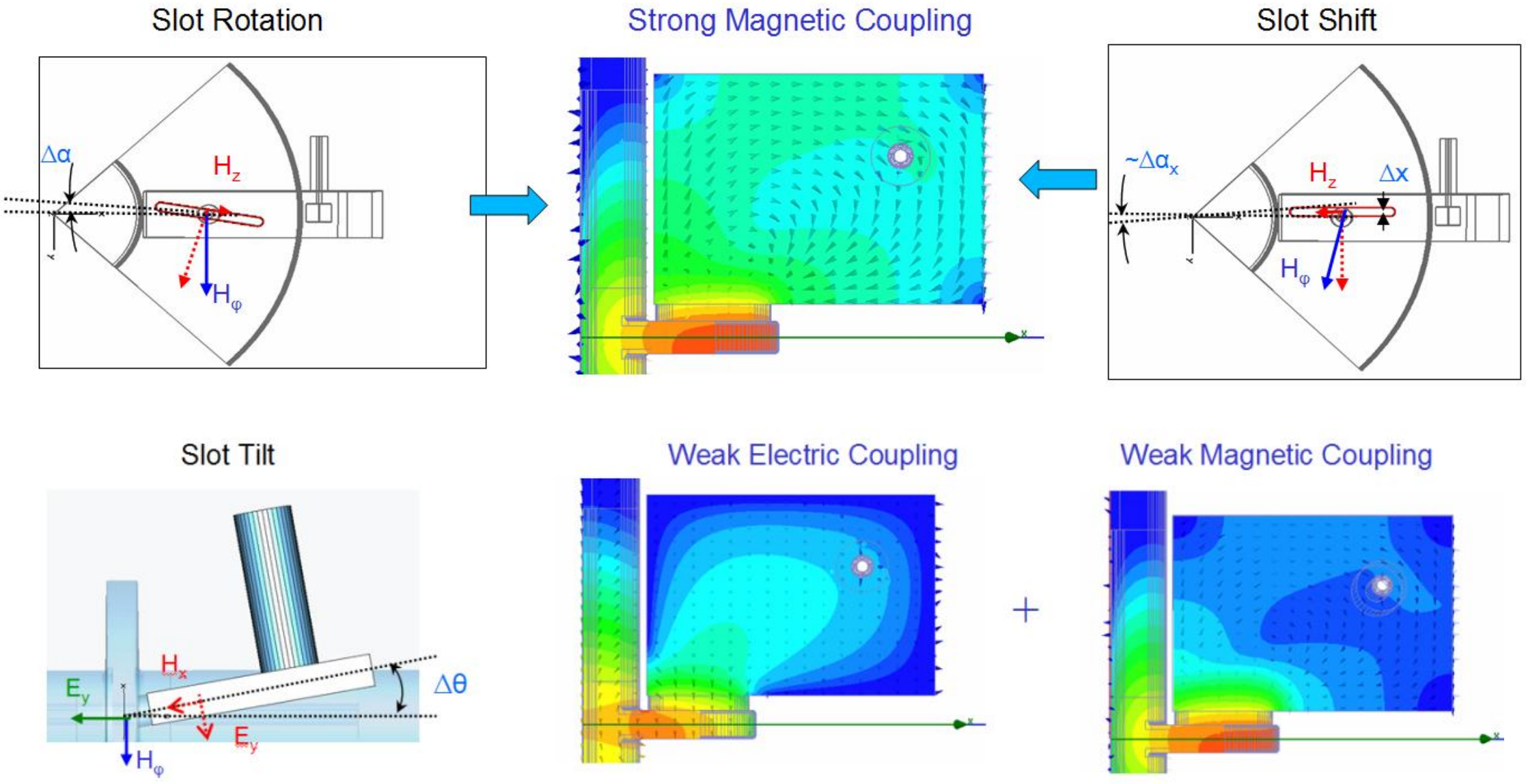

Figure 5: Investigation of tolerances on the cavity-waveguide coupling slot.

## SPATIAL RESOLUTION LIMITS

Table 3: Limitations of the BPM resolution due to $TM_{010}$ & $TM_{210}$ mode leakage.

| Mode Type | Freq. [GHz] | $Q_{tot}$[1] | Beam shift [μm] | Output voltage[2] [mV] | BPM Resolution [nm] | |
|---|---|---|---|---|---|---|
| | | | | | SB | MB |
| $TM_{010}$ | 10.385 | 380 | 0 | <1 | 40 | 4 |
| $TM_{110}$ | 13.999 | 250 | 0.1 | 2.4 | - | - |
| $TM_{210}$ | 18.465 | 80 | 100 | <0.18 | 8 | 1 |
| $TM_{210}$ | 18.465 | 80 | 500 | <4 | 200 | 20 |

1 – Stainless steel material was used.
2 – RMS value of the sum signal of two opposite coaxial ports at the 14 GHz operating frequency after all filters applied; signals are normalized to 1 nC charge

The spatial resolution of the cavity BPM is defined as the smallest change of the beam position, which can be resolved. Beside the mode leakage effects, we also have to investigate consequences of practical imperfections, like mechanical tolerances from the manufacturing process, and other imperfections. It turns out, that the alignment of the coupling slot windows between cavity resonator and waveguides are most critical. In detailed EM-simulations we studied the effects of asymmetries due to shift, rotation and tilt of these slots, which causes additional leakage of the higher-order modes (see Figure 5). Also a shift between beam pipe and cavity centers will cause a degradation of the spatial resolution. In summary, limiting the slot shift to <5 μm, and the slot rotation to <0.05 deg, we can expect the cross coupling between horizontal and vertical plane to be >40 dB, while achieving a ±100 μm dynamic range at maximum resolution. This resolution is dominated by the $TM_{010}$ mode leakage, and is ~40 nm for a single bunch (SB) excitation, and ~4 nm for a multi-bunch (MB) beam batch (see Table 3). At large beam displacements >0.5 mm, the $TM_{210}$ mode leakage dominates and limits the theoretical achievable resolution, however, in practice the limited dynamic range of the read-out receiver will further reduce the BPM resolution for this case.

## CAVITY BPM PROTOTYPE FOR CTF

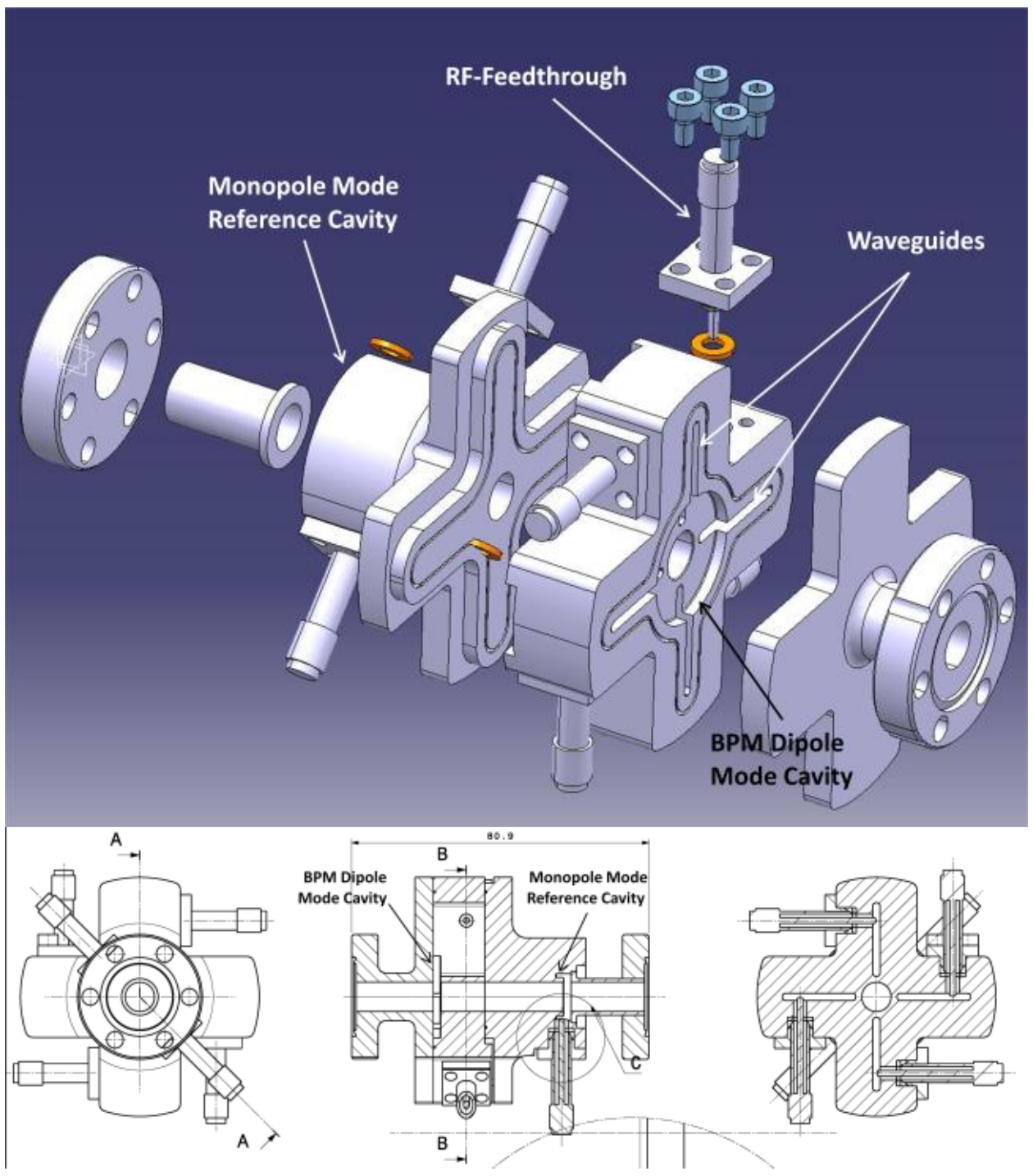

Figure 6: Layout of the CTF cavity BPM prototype.

Figure 6 shows an early draft of the mechanical layout of the CTF cavity BPM prototype. It consists out of four major parts, all milled / EDM out of type 304 stainless steel:

- The central section with the BPM resonator and waveguides. The waveguide structures can be manufactured using a wire-EDM process, the cavity will be milled.
- A section with the reference cavity and a cover for the waveguides
- Two cover sections with transitions for the beam pipes

The manufacturing tolerances on the critical structures have to be <±5 µm. We will use four RF-feedthroughs with capacitive pin coupling for the waveguide-coaxial transition of the BPM signals, and additional two RF-feedthroughs with magnetic coupling, extracting the signal of the reference cavity. The tolerances of the capacitive pin coupling are very tight, for the prototype we consider tuning dimples. We plan to use a commercial microwave UHV-feedthrough with minor custom modifications (*Meggitt*); some details still have to be worked out.

The total length of the cavity BPM detector including flanges measures 80.9 mm, BPM and reference cavities are separated ~50 mm. Both resonators operate at the same frequency (~15 GHz) and have very similar loaded Q-values (~300). The signal from the reference cavity can be used directly (no external frequency or clock signals required), and provides

- A phase reference for the dipole mode signal, to distinguish the beam position is up or down, respectively left or right.
- A magnitude reference, as the dipole mode signal is proportional to beam displacement and intensity.

A set of three CTF prototype cavity BPMs have to be manufactured, tested and installed into the CTF beam-line to fully exploit beam based studies on the BPM characteristics and measurement capabilities. The presented cavity BPM detector has to be complemented by a read-out system, utilizing microwave and digital signal processing technologies.

## SUMMARY

A simple, straightforward design of a cavity BPM with high spatial (50 nm) and temporal (50 nsec) resolution is proposed for the CLIC Main Linac. In depth EM-simulations and optimizations include the analysis of mechanical tolerances. A prototype with a slightly different operation frequency, to be tested at the CLIC Test Facility (CTF), is currently under development.